\documentclass[twocolumn,pre,aps,showpacs,superscriptaddress]{revtex4}
\usepackage{amsmath}
\usepackage[psamsfonts]{amssymb}
\usepackage{bm}
\usepackage{graphicx}
\usepackage{dcolumn}
\usepackage{color}
\usepackage{natbib}

\def\beq{\begin{equation}}
\def\eeq{\end{equation}}
\def\bea{\begin{eqnarray}}
\def\eea{\end{eqnarray}}

\newcommand{\vth}{\ensuremath{v_\text{th}}}
\newcommand{\nubar}{\ensuremath{\bar{\nu}_\text{hc}}}
\newcommand{\nusc}{\ensuremath{\nu_\text{hc}}}

\begin{document}

\title{Quantum statistical model of nonlinear inverse bremsstrahlung absorption
       \\in strongly coupled plasmas}

\author{A.\ Grinenko}
\email{A.Greenenko@Warwick.ac.uk}
\affiliation{Centre for Fusion, Space and Astrophysics,
             Department of Physics, University of Warwick,
             Coventry CV4 7AL, United Kingdom}
\author{D.O.\ Gericke}
\affiliation{Centre for Fusion, Space and Astrophysics,
             Department of Physics, University of Warwick,
             Coventry CV4 7AL, United Kingdom}

\date{\today}

\begin{abstract}
A new approach for the calculation of collisional inverse bremsstrahlung
absorption of laser light in dense plasmas is presented. Quantum statistical
formalism used allows avoiding {\em ad hoc} cutoffs that were necessary in
classical approaches. Thus, the current method remains reliable for strong
electron-ion interactions. In addition, both the dynamic, field dependent
response and hard electron-ion collisions, are consistently incorporated. The
latter were treated in an average manner as a stopping power that in turn was
cast into a form of a friction force. Here, for the first time a link between
the stopping power and the problem of collisional laser absorption is drawn.
This allows the theories developed for the stopping power calculation, such as
the quantum T-matrix approach, to be applied to the problem of collisional laser
absorption. The new approach accommodates the low- and high-frequency limits
explained in the text and is valid for arbitrary laser field intensities. A
comparison with classical MD simulation is indicative of the validity of the
new method in the wide parameter range tested.
\end{abstract}

\maketitle

\section{Introduction}
Understanding the laser-matter interaction in strongly coupled plasmas is
crucial for the design of the contemporary inertial confinement fusion
(ICF) targets. In both the direct and indirect drive ignition schemes, the laser
energy is deposited into the plasma of high-Z elements such as Au. The
conditions imposed by the hydrodynamic instabilities on the spatial symmetry of
the laser energy deposition, require the laser absorption to be carefully
determined from the early stages of the laser-plasma interaction
\cite{Lindl2004}. In the case of the fast ignition \cite{Key2007} it is
important to model the laser absorption starting with the preformed plasma due
to the unavoidable nanosecond prepulse. The critical plasma density
corresponding to the third harmonic of the Ny:Yag lasers, used in most inertial
fusion designs is $\sim 10^{22}$~cm$^{-3}$. In the limit of low temperatures
(few eV) such plasmas are characterised by strongly coupled and degenerate
electrons: $\Gamma \!=\! (e^2/k_B T_e) (4\pi n_e/3)^{1/3} \sim 1$ and 
$n_e (2\pi\hbar^2/m_e k_B T_e)^3 \sim 1$ respectively. 

The dominant mechanism of radiation absorption for lasers with intensities
typical for ICF is the inverse bremsstrahlung. In this case, the radiation
is absorbed via collisions between the plasma particles usually described in
terms of the electron-ion collision frequency \cite{Silin1965}. First
calculation of the inverse bremsstrahlung in the high frequency limit for the
lowest order in the laser field strength was reported by Dawson \& Oberman (DO)
\cite{Dawson1962}. Later Decker {\em et al.} \cite{Decker1994} have extended
this result to arbitrary field strengths. A classical ballistic model was
considered in Ref.~\onlinecite{Mulser2001} in order to study the
frequency-dependent electron-ion collisions in plasmas. 

These approaches are formulated in the high-frequency limit, when the number of
binary electron-ion collisions per laser cycle can be neglected. In this limit,
the laser is coupled to the plasma via the induced polarisation current so
that the electron-ion interaction has a collective rather than a binary
character. In the low-frequency limit strong scattering due to the binary
collisions dominate the process of the absorption and the induced polarisation
becomes relatively small. At intermidiate frequencies the effects of binary and
collective scattering have to be considered simultaneously. 

For a given laser frequency, plasma conditions can be such, that the rate of
the binary collisions becomes comparable with the laser period. Considering
$0.35$ $\mu$m lasers used in most applications, this occurs in a strongly
coupled plasmas. The collisions in such plasmas have to be evaluated quantum
mechanically. The first quantum  treatment was reported in
Ref.~\onlinecite{Perel'1962}. The nonlinear absorption was determined using the
first Born approximation in Refs. \cite{Rand1964,Schlessinger1979}. A quantum
approach to calculate laser absorption in strong fields was also developed in
Ref.~\onlinecite{Silin1981}. Semiclassical approach in the linear regime using a
memory function kinetic formalism including lowest order quantum effects was
developed in Ref. \cite{Cauble1985}. A quantum statistical approach for
dynamical conductivity in strongly coupled regime was developed in Refs.
\cite{Ropke1998,Reinholz2000}. A quantum Vlasov approach for arbitrary field
strength, similar to the classical approach developed by Decker {\em et al.}
\cite{Decker1994} was presented in Ref.~\onlinecite{Kull2001}. Rigorous kinetic
approach to the inverse bremsstrahlung absorption in strongly coupled plasmas
using nonequilibrium Green's function techniques was developed in a series of
publications \cite{Kremp1999,Bornath2001}. The expression for the collision
frequency derived by \citeauthor{Bornath2001} using the latter approach is
identical to the quantum Vlasov method \cite{Kull2001}. In fact, it can be
obtained from the formula obtained by \citeauthor{Decker1994} by replacing the
classical dielectric function by the quantum Lindhard dielectric function
\cite{Lindhard1954,Ashcroft1976}. This indicates, that both approaches make
similar approximation, neglecting the effect of strong scattering by binary
collisions, which makes them applicable in the high-frequency limit only. The
effect of the binary collisions was considered in a linear-response theory by
using the Gould-DeWitt scheme in Ref.~\onlinecite{Wierling2001}. Inverse
bresstrahlung absorption for strongly coupled plasmas was also calculated using
classical molecular dynamic (MD) simulations reported in
Refs.~\cite{Pfalzner1998,Hilse2005,Morozov2005}.

In this paper, we present a description of collisional absorption that bridges
between the high- and low-frequency limits. It is shown that the interactions
can be split into a weak collective interactions and hard collisions. The latter
are treated as the stopping power of ions in the electron fluid that can be cast
into the form of a friction between the electron and ion fluids. Thus, for the
first time, the stopping power formalism is applied to the calculation of the
collisional absorption, allowing one to use a well developed models of the
stopping power (see, e.g., Refs.~\cite{Gericke1999,Zwicknagel1999,Gericke2002})
in the problem of laser absorption in plasmas. The description of the collective
electron response can be kept almost unchanged from earlier approaches
\cite{Decker1994,Kull2001}. Due to the use of full quantum mechanical
formulation of the problem, no {\em ad hoc} cutoffs must be introduced and the
theory stays reliable for strong electron-ion interactions and degenerate
electrons. The few assumptions made are justified by the unprecedented agreement
with molecular dynamic (MD) simulations
\cite{Pfalzner1998,Hilse2005,Morozov2005} up to very high coupling strengths.

In the following chapters we develop a quantum formalism to describe the laser
absorption in dense plasmas. In Sec.~\ref{sec:1} a set of Vlasov-Poisson (VP)
equations similar to the one developed by \citeauthor{Kull2001} is introduced.
The major improvement in the present model is the inclusion of the
hard-collisions and its treatment using the stopping power formalism. In
Sec.~\ref{sec:2} the VP equations are solved in the Kramers-Henneberger
(KH) frame that is calculated with respect to the effect of the hard-collisions
on the electron fluid rest-frame. Results and discussion follow in
Sec.~\ref{sec:3}.

\section{Quantum Vlasov Equation with an Effective Friction Force \label{sec:1}}
The motion of the electrons in a neutral plasma consisting of $N_i$
ions and $N_e = Z N_i$ electrons, where $Z$ is the average charge state is
described by a one electron statistical operator $\widehat{\rho}_1(t)$ whose
evolution is governed by equation:
\beq
i\hbar\frac{\partial}{\partial t}
 \widehat{\rho}_1-\left[\widehat{H}_0,\widehat{\rho}_1\right ]
 = \mathop{\rm Tr}_{2}\{N_e \widehat{V}_{1,2}^{ee} \widehat{\rho}_{1,2}^{ee}
                  +N_i \widehat{V}_{1,2}^{ei}\widehat{\rho}_{1,2}^{ei}\},
\label{eq:rho1_general}
\eeq
The ions are considered to be located at fixed positions $\widehat{\bm x}_j$,
$j=1,2,\dots,N_i$ distributed according to the temperature dependant ion-ion
pair correlation function $g_{ii}(\bm x)$. The Hamiltonian of the system is 
\beq
\widehat{H}_0=\frac{\bm{\widehat{p}}_1^2}{2m}+\widehat{U}_{ext},
\eeq
where $\widehat{U}_{ext}$ describes the externally applied field,
$\widehat{\rho}_{1,2}^{ee}$ and $\widehat{\rho}_{1,2}^{ei}$ are the two particle
electron-electron and electron-ion distribution functions respectively,
$\widehat{V}_{1,2}^{ee}$ and $\widehat{V}_{1,2}^{ei}$ are the e-e and e-i
interaction potentials. Next assuming that the two-particle density functions
can be expressed as:
\beq
 \widehat{\rho}_{1,2}^{\alpha\beta} =
 \widehat{\rho}_{1}^{\alpha}\widehat{\rho}_{2}^{\beta}+
 \widehat{g}_{1,2}^{\alpha\beta}
 \label{eq:pair_correlation}
\eeq
The first term in the expansion of $\widehat{\rho}_{1,2}^{\alpha\beta}$ on the
r.h.s is the Hartree term, while all the higher order terms are contained in the
the corresponding pair correlation function $\widehat{g}_{1,2}^{\alpha\beta}$.
Using these definitions Eq.~\eqref{eq:rho1_general} can be written as:
\beq
i\hbar\frac{\partial}{\partial t}
 \widehat{\rho}_1-\left[\widehat{H},\widehat{\rho}_1\right ]
 = \mathop{\rm Tr}_{2}\{N_e \widehat{V}_{1,2}^{ee} \widehat{g}_{1,2}^{ee}
                  +N_i \widehat{V}_{1,2}^{ei}\widehat{g}_{1,2}^{ei}\},
\label{eq:rho1_boltzmann}
\eeq
where 
\beq
\widehat{H}=\widehat{H}_0-e\widehat{\Phi},
\label{eq:Hamiltonian}
\eeq
and the effective Hartree potential $\Phi$ is:
\beq
-e\widehat{\Phi} = \mathop{\rm Tr}_{2}\{N_e \widehat{V}_{1,2}^{ee}
\widehat{\rho}_{2}^{e}
                  +N_i \widehat{V}_{1,2}^{ei}\widehat{\rho}_{2}^{i}\}
\eeq
In Eq.~\eqref{eq:rho1_boltzmann} the hard, binary, collisions are grouped in the
r.h.s. and weak collisions are treated in the framework of Hartree approximation
as a collective average potential contributed by the system species. This
effective potential is determined self-consistently from the charge densities of
electrons and ions by the Poisson equation:
\beq
\vartriangle \Phi(\bm x, t) = 4\pi e \Bigl( 
n_e(\bm x, t) -Z\sum_{j=1}^{Ni}\delta(\bm x-\bm x_j)
\Bigr)
\label{eq:Poisson}
\eeq
where $n_e(\bm x, t) = N_e \langle\bm x|\widehat{\rho}_{1}|\bm x\rangle$ is
the electron density. The potential is calculated as a classical field
$\Phi(\bm x, t)$ which is used as an operator $\Phi(\widehat{\bm x}, t)$ in
Eq.~\eqref{eq:rho1_boltzmann}. Thus, assuming
electrostatic interactions with a self consistent collective scattering
potential $\Phi(\bm x, t)$ and an externally applied potential
$\Phi_{ext} = -{\bm x}\cdot{\bm E}_{ext}(t)$, due to a time dependent laser
field ${\bm E}_{ext}(t) = {\bm E}_0 \sin(\omega_0 t)$, the Hamiltonian can be
written as
\beq
 \widehat{H} = \frac{\widehat{p}_1^2}{2m}
 -e\left[\Phi(\widehat{\bm x}_1,t)+
  \Phi_{ext}(\widehat{\bm x}_1,t)\right]
\label{eq:H}
\eeq

The heating rate of the plasma by the external laser field can be expressed
using the effective electron-ion collision frequency \cite{Silin1965} given by
the following expression:
\beq
\nu_{ei} = \frac{4\pi \omega_0^2}{\omega_p^2}
\frac{\overline{\langle \widehat{\bm j} \rangle\cdot {\bm E} }}
{ \overline{{\bm E} \cdot {\bm E}} }
\equiv \frac{4\pi \omega_0^2}{\omega_p^2}\sigma.
\label{eq:Silin}
\eeq
Here, the overline stands for averaging over one oscillation period and the
angular brackets denote the expectation values of the quantum operators. In
the last equation $\widehat{\bm j}$ is the electric current density operator:
$\widehat{\bm j} = -(e n_e/m) \widehat{\bm p}_1$. In order to determine the time
dependence of the expectation values of the operators, one has to switch from
the microscopic quantities given by the kinetic equations to the
average macroscopic quantities via the statistical operators. Thus, multiplying
by $\widehat{\bm p}_1$ and applying the trace to the both sides of
Eq.~\eqref{eq:rho1_boltzmann}, the time change-rate of the expectation value of
the momentum $\langle\widehat{\bm p}_1\rangle$ is obtained 
\beq
i\hbar\frac{\partial}{\partial t}\langle\widehat{\bm p}_1\rangle -
\left\langle\left[\widehat{\bm p}_1,\widehat{H}\right]\right\rangle =
i\hbar\mathop{\rm Tr}_{1}{\widehat{I}_{1}^{hc}\widehat{\bm p}_1},
\label{eq:mot_momentum}
\eeq
where the hard-collision integral $\widehat{I}_{1}^{hc}$ is defined by:
\beq
 \widehat{I}_{1}^{hc} \equiv 
 \mathop{\rm Tr}_{2}\{N_i\widehat{V}_{1,2}^{ei}\widehat{g}_{1,2}^{ei}+
                      N_e\widehat{V}_{1,2}^{ee}\widehat{g}_{1,2}^{ee}
                    \}/i\hbar.
\eeq
The integral on the right-hand side contains hard-collisions only and is related
to the hard-collisions contribution to the stopping power by
\beq
\left\langle\frac{dE}{dx}\right\rangle=\mathop{\rm Tr}_{1}
\bigl\{\widehat{I}_{1}^{hc}(t)\bm{\widehat{p}}_1\bigr\}.
\eeq
In order to obtain an analytical solution of Eq.~\eqref{eq:Silin}, it is
useful to cast the effect of the stopping power as an average friction force
between the electron and ion fluids $R \equiv -\langle dE/dx \rangle/V$, where 
$\bm V = \langle \widehat{\bm p}_1\rangle/m$ is the average ensemble velocity,
and restrict the solution to small particle velocities, such that $V
\lesssim \vth$, where $\vth \equiv kT_e/m$ is the electron thermal velocity,
since in this case the friction coefficient is velocity independent
\cite{Gericke1999,Gericke2003}. The dynamics of screening is of minor importance
in the low velocity range, justifying the use of the stopping power data
calculated assuming statically screened Coulomb interactions
\cite{Gericke1999,Gericke2003}. In this limit, the rate equation for the average
momentum becomes
\beq
\frac{d}{dt}\langle\widehat{\bm p}_1\rangle = 
-e\bigl[
{\bm E_0} \sin{(\omega_0 t)}+ \langle \widehat{\bm {E}} \rangle  
\bigr]
-\nusc \langle\widehat{\bm p}_1\rangle,
\label{eq:mot_current}
\eeq
where the hard collision frequency $\nusc$ is defined as $\nusc\equiv R/m$ and
the polarisation field is $\langle\widehat{\bm E} \rangle \equiv \langle
{d\widehat{\Phi}}/d\widehat{\bm x} \rangle$. 

Field amplitudes for which the assumption of velocity independent friction
coefficient $R$ is valid are restricted by the condition $V_{\rm max}<\vth$.
Neglecting the contribution of the polarisation field $\langle \widehat{\bm E}
\rangle$ in Eq.~\eqref{eq:mot_current}, the maximum average velocity can be
estimated by
\beq
V_\text{max} = \frac{{1+\nubar}}{{1+\nubar^2}}v_0, 
\label{eq:V_max}
\eeq
where $v_0 \equiv e E_0 / m\omega_0$ is the free electron quiver velocity, and
$\nubar \equiv \nusc/\omega_0 $ is the normalised hard-collision frequency. 
Therefore, in the high-frequency limit: $\nubar \ll 1$ the fields amplitudes 
are restricted by $v_0 \lesssim \vth$; and in the low-frequency limit: $\nubar
\gg 1$ the condition becomes $v_0 \lesssim \nubar\,\vth  $. Thus, because of
the strong damping of the velocity in the low-frequency limit, the region where
$R={\rm const}$ is applicable can be extended to stronger fields. 

Furthermore, the use of the stopping power formalism imposes additional 
limitation with respect to the laser frequency $\omega_0$. The stopping power
treatment assumes that the typical collision time is greater than the period of
plasma oscillation characterised by the plasma frequency
$\omega_p\equiv\sqrt{4\pi e^2n_e/m}$. Therefore, the analytical solution to be
obtained is restricted to the laser frequency range of $\omega_0\lesssim
\omega_p$.

Multiplying both sides of Eq.~\eqref{eq:mot_current} by $-e n_e/m$ the current
balance equation is obtained 
\beq
\frac{d\langle \widehat{\bm j} \rangle}{dt} = \frac{\omega_p^2}{4\pi}
\bigl[
{\bm E_0} \sin{(\omega_0 t)}+ \langle \widehat{\bm E} \rangle
\bigr]
-\nusc \langle\widehat{\bm j}\rangle
\label{eq:j_mot_current}
\eeq
Note, that in the low-frequency limit $\nubar \gg 1$, the polarisation field
$\langle\widehat{\bm E} \rangle$ vanishes and Eq.~\eqref{eq:mot_current}
produces the well known Drude formula for the low-frequency conductivity:
\beq
4\pi\sigma_D = \frac{\omega_p^2}{\nusc-i\omega_0}. 
\eeq

A formal solution of
Eq.~\eqref{eq:j_mot_current} is
\beq
\begin{split}
\langle \widehat{\bm j} \rangle (t)  =
&-\frac{\omega_p^2}{4\pi}\frac{\gamma {\bm E_0}}{\omega_0} 
	\left[
		\cos{(\omega_0t)}-\nubar\sin{(\omega_0t)}
	\right]+ \\ &
	\frac{\omega_p^2}{4\pi} 
	\int\limits_{-\infty}^{t} 
	\langle\widehat{\bm E}\rangle(\tau) e^{\nusc (\tau-t)} d\tau
\label{eq:current}
\end{split}
\eeq
where $\gamma \equiv {1}/(1+\nubar^2)$. One can see that the current consists
of two terms, namely the polarisation current represented by the second term on
the r.h.s., and the free current, represented by the first term. The latter
includes a contribution from the hard collisions that are in phase with the
laser field. The collision rate is obtained by substituting
Eq.~\eqref{eq:current} into Eq.~\eqref{eq:Silin}:
\beq
\begin{split}
\nu_{ei}  & \!=\! \gamma\nusc\!+\! 
\frac{2\omega_0^2}{E_0^2}
\overline{\left (
{\bm E_0} \sin{(\omega_0t)} \!\!
\int\limits_{-\infty}^{t}\!\!
\langle \widehat{\bm E}\rangle(\tau)
e^{\nusc (\tau-t)} d\tau 
\right ) }
\label{eq:frequency}
\end{split}
\eeq
Here, the first term is due to the the strong collisions. It has the same
form as the real part of the Drude conductivity. The second term is due to the
polarisation current. 

To complete the derivation one needs to calculate the polarisation field
$\langle \widehat{\bm E}\rangle$ to be used in the last equation. To do so the
system of Eq.~\eqref{eq:rho1_boltzmann} and Eq.~\eqref{eq:Poisson} has to be
solved. The general solution of this problem is notoriously difficult however,
it can be greatly simplified if one assumes that the same average friction force
$-\nusc\langle\widehat{\bm p}_1\rangle$ is acting on all the electrons
irrespective of their direction, position and velocity. This assumption is
justified in the limit of  $V_{\rm max} \gg \vth^{\rm ion}$, when the ions
directed motion as a particle beam characterised by a single velocity with
respect to the electrons can be considered. In such case,
Eq.~\eqref{eq:rho1_boltzmann} can be cast into the form of the quantum Vlasov
equation:
\beq
i\hbar\frac{\partial}{\partial t}
 \widehat{\rho}_1-\left[\widehat{H}_{\rm eff},\widehat{\rho}_1\right ] = 0,
\label{eq:vlasov}
\eeq
where the effective Hamiltonian $\widehat{H}_{\rm eff}$ is:
\beq
\widehat{H}_{\rm eff} \equiv
\frac{\widehat{p}_1^2}{2m}-e\left(\widehat{\Phi}+
	\widehat{\Phi}_{\rm ext}\right)
	+\nusc\langle\widehat{\bm p}_1\rangle\widehat{\bm x}_1
\eeq
Eq.~\eqref{eq:vlasov} produces the same current balance equation
(\ref{eq:j_mot_current}) as Eq.~\eqref{eq:rho1_boltzmann}. The set of two
equations: Eq.~\eqref{eq:Poisson} an Eq.~\eqref{eq:vlasov} form a closed set of
Vlasov-Poisson (VP) equations with the hard collision determined using the
standard methods applied for the stopping power calculations.

\section{Solution of quantum VP equations \label{sec:2}}
In this chapter we shall obtain the solution of the system of VP equations
introduced above. The laser and the friction act both as an effective external
force since they depend on $\hat{\bm x}$ only. Therefore, in the absence of the
scattering field $\widehat{\Phi}$ the electrons would perform a quiver motion.
It is useful to transform to Kramers-Henneberger (KH) reference frame --
the rest frame of the electron fluid, since one can assume that in this frame
the electrons are close to the equilibrium unperturbed state. The transformation
to KH frame is given by $\bm{y=x-\xi}$, and
\beq
{\bm \xi} =  -\gamma {\bm \epsilon}
\left[
\sin {(\omega_0 t)}+\nubar\cos {(\omega_0 t)}
\right] 
\label{eq:transformation}
\eeq
where ${\bm \epsilon} = -e{\bm E_0}/m\omega_0^2$. In this frame the VP
equations become:
\begin{subequations}
\beq
i\hbar\frac{\partial}{\partial t}
 \widehat{\rho}^{\rm KH}-\left[\frac{\widehat{p}^2}{2m}
                        -e\widehat{\Phi}^{\rm KH},
			\widehat{\rho}^{\rm KH}\right ] = 0
\eeq
\beq
\vartriangle \Phi^{\rm KH}(\bm x, t) = 4\pi e 
\Bigl( 
n_e(\bm y, t) -Z\sum_{j=1}^{Ni}\delta(\bm y+\bm x_j)
\Bigr) \label{eq:VP_KH_Phi}
\eeq
\label{eq:VP_KH}
\end{subequations}
Equations \eqref{eq:transformation} and \eqref{eq:VP_KH} form the basis
for the calculation of the laser energy absorption by the electron-ion
scattering. Since in the following analysis we shall carry out the calculation
mainly in the KH frame, we shall omit the KH superscript. The solution procedure
of this set of equations in identical to the solution procedure of the VP
equations (4a) and (4b) discussed in Ref.~\onlinecite{Kull2001}. The only
difference is in the definition of the KH frame. Here, the effect of the hard
collisions in the form of the friction force is included in the determination
of the KH frame, whereas in Ref.~\onlinecite{Kull2001} the electrons are freely
oscillating in the external field. Therefore, we adopt the notations used in
Ref.~\onlinecite{Kull2001} and skip some parts of the derivation elaborated in
Ref.~\onlinecite{Kull2001} while underlining the differences obtained in the end
result. 

In the KH frame, the scatterring potential is treated as a perturbation
$\widehat{H}^{(1)} = -e\widehat{\Phi}_{\rm eff}$ of the equilibrium set up by
the Hamiltonian $\widehat{H}^{(0)} = \widehat{p}^2/2m$. Correspondingly, the
density operator is set to be
$\widehat{\rho}=\widehat{\rho}^{(0)}+\widehat{\rho}^{(1)}$, where
$\widehat{\rho}^{(0)}$ is the unperturbed stationary ensemble.

The Fourier-Laplace transform of the effective potential
Eq.~\eqref{eq:VP_KH_Phi} is found to be:
\beq
\Phi(\bm k, \omega) = \frac{\Sigma_i(\bm k, \omega)}{D(\bm k, \omega)}
\eeq
where $\Sigma_i(\bm k, \omega)$ is the Fourier-Laplace transform of the ion
potential in the KH frame. The Fourier transform assumes the form,
\beq
\Sigma_i(\bm k, t)=\Sigma_{i,0}(\bm k)\times e^{i\bm k\cdot \bm \xi(t)}
\label{eq:sigma_kt}
\eeq
where
\beq
\Sigma_{i,0}(\bm k)=\frac{4\pi Z e}{k^2}\sum_j e^{-i\bm k \cdot \bm x_j}
\eeq
is a static part and
\beq
e^{i\bm k\cdot \bm \xi(t)} = 
                   \sum_{n,m}^{\infty} 
		   (-1)^n (-i)^m J_n(\gamma z)J_m(\gamma\nubar z)
		   e^{i(n+m)\omega_0 t}, \nonumber
\eeq
is a dynamic phase factor due to the quiver motion. Its Fourier
coefficients are Bessel functions of the first kind $J_n$ depending on the
parameter $z\equiv\bm k \cdot \bm \epsilon$. Applying the Laplace
transformation to Eq.~\eqref{eq:sigma_kt} yields
\beq
\Sigma_i({\bm k}, \omega) = \Sigma_{i,0}(\bm k)
     \sum_{n,m}^{\infty}
     \frac{(-1)^n(-i)^m
     J_n\left( \gamma z \right)
     J_m\left( \gamma\nubar\,z \right)}
     {-i\bigl(\omega+(n+m)\omega_0\bigr)}
\eeq
The electric potential generated by the ions in the KH frame is screened by
the dielectric function $D({\bm k},\omega)$ 
\beq
D({\bm k},\omega) = 1+4\pi\chi(\bm k, \omega)
\eeq
which is known as the Lindhard dielectric function
\cite{Lindhard1954,Ashcroft1976} and the response function $\chi(\bm k,\omega)$
is given by
\beq
\chi({\bm k},\omega) = \frac{e^2}{k^2}
\int d^3 {\bm u}
\frac{f(\bm u +\hbar \bm k/2m)-f(\bm u -\hbar \bm k/2m)}
{\hbar(\omega-{\bm k}\cdot {\bm u})}
\label{eq:dielectric}
\eeq
The perturbed charge density 
\beq
\tau_e(\bm k, \omega) \equiv -eN_e\rho^{(1)}(\bm k, \omega) \nonumber
\eeq
is related to the effective potential and the response function by
\beq
\tau_e(\bm k,\omega)=-k^2\chi(\bm k, \omega)\Phi(\bm k,\omega) 
\eeq
Finally, we apply the inverse Laplace transform to  $\tau_e(\bm k,\omega)$ and
$\Phi(\bm k,\omega)$ and obtain from the poles of $\Sigma_i(\bm k,\omega)$ at
the frequencies $\omega = -(n+m)\omega_0$ the asymptotic result,
\begin{subequations}
 \beq
  \Phi(\bm k,t) = \sum_{n,m}^{\infty} \Phi_{n,m}(\bm k)e^{i(n+m)\omega_0 t} \\
 \eeq
 \beq
  \Phi(\bm k,t) = \sum_{n,m}^{\infty} \Phi_{n,m}(\bm k)e^{i(n+m)\omega_0 t} \\
 \eeq
 \label{eq:phi_tau}
\end{subequations}
where the coefficients of these series are given by
\begin{subequations}
 \beq
  \Phi_{n,m}(\bm k) = \Sigma_{i,0}(\bm k)
     \sum_{n,m}^{\infty}
     \frac{(-1)^n(-i)^m
     J_n\left( \gamma z \right)
     J_m\left( \gamma\nubar\,z \right)}
     {D\bigl(\bm k,-(n+m)\omega_0\bigr)}
 \eeq
 \beq
  \tau_{n,m}(\bm k) = -k^2\chi\bigl(\bm k,-(n+m)\omega_0\bigr)\Phi_{n,m}(\bm k)
 \eeq
\end{subequations}
These results allow us to obtain the expectation value of the electric field
used in the calculation of the collision frequency in Eq.~\eqref{eq:frequency}.
In the momentum representation $\bm E (\bm k, t)=-i\bm k \Phi(\bm k, t)$ and
using the transformations (A8b), (A15) from Ref.~\onlinecite{Kull2001} yields
\beq
 \langle \bm E(\bm k, t) \rangle = -\frac{1}{e N_e}\int 
 \frac{d^3\bm k}{(2\pi)^3} i\bm k \Phi^*(\bm k, t)\tau_e(\bm k, t)
\eeq
Substituting this expression into Eq.~\eqref{eq:frequency} and using
Eqs.~\eqref{eq:phi_tau} we get the collision frequency
\beq
\begin{split}
\label{eq:coll_freq}
 &\frac{ \nu_{ei} }{\omega_p}  = 
 \frac{\gamma\nusc}{\omega_p}-
 \!\!\!{\sum_{\substack{m=-\infty\\n=-\infty}}^{\infty}} \alpha
 \int\limits_0^\infty d^3 {\bm k}\, \frac{i}{k^2}
 \frac{
 J_n\bigl( \gamma z\bigr)
 J_m\bigl( \gamma \nubar z \bigr)
 }
 {
 D\bigl({\bm k},(n+m)\omega_0\bigr)
 } \\&
	\sum_{s=-\infty}^{\infty} i^s 
		J_{m+s}\bigl( \gamma \nubar z \bigr)	
		\Bigl [
		(n-s)(1+i\,\nubar)
		J_{n-s}\left( \gamma z \right)\\&
		-i\,(\gamma\nubar\, z) 
		J_{n-s-1}\left( \gamma z \right)
		\Bigr ]S_\text{ii}(\bm k)
\end{split} 
\eeq
where, $S_\text{ii}(\bm k)$ is the ion-ion structure factor, and 
\beq
\alpha \equiv
(2\pi^2)^{-1}(\omega_0/\omega_p)(Ze^2/m v_0^2)
\eeq
Clearly, the known limiting cases can be readily retrieved. Decker's result (Eq.
20 in Ref.~\onlinecite{Decker1994}) follows from 
Eq.~\eqref{eq:coll_freq} in the weak coupling limit $\nubar \!\to\! 0$ and
non-degenerate plasmas. In this case $D(\bm k,\omega)$ becomes the classical
dielectric function and the integral must be truncated at $k_{max}$ to avoid the
divergence at small impact parameters (see Ref.~\onlinecite{Grinenko2009} for
the discussion of different cuttofs). Here, all integrals can be performed to
infinity and no {\em ad hoc} cutoffs must be introduced as a result of the
quantum mechanical treatmen. The first term in line 1 dominates for
small laser frequencies $\nubar\!\to\!\infty$ giving a Drude-like
expression.

\section{Results and Discussion \label{sec:3}}
In the previous sections we have presented a quantum mechanical formulation for
the problem of laser absorption in dense plasmas. The formalism is similar to
that developed by \citeauthor{Kull2001}, but inherently includes the
hard-collisions absent from this and other approaches
\cite{Decker1994,Kremp1999,Bornath2001}. This was achieved in three steps.
Firstly, the the two-particle density function was split into: (i) the first
order Hartree term, representing the weak interactions; (ii) the higher
orders, representing the hard-collisions, which were collected into the pair
correlation function (e.g. Eq.~\eqref{eq:pair_correlation}). Secondly, the
hard-collisions were cast into the form of the friction force using the
stopping-power integral on the right-hand side of the momentum rate equation
\eqref{eq:mot_momentum}. At last, the collision frequency Eq.~\eqref{eq:Silin}
was determined by finding the first order perturbation of the equilibrium
density distribution as a result of the electron-ion scattering. The equilibrium
electron distribution is set up in the electron KH rest frame determined by the
external potential and the friction force due to the hard-collisions.

The use of the stopping power formalism restricts this approach by demanding
that: (i) the interaction time is larger than or at least of the same order of
magnitude as the typical plasma oscillation period, setting
$\omega_0\lesssim\omega_p$; (ii) the field strength is sufficiently high to
consider the ion motion relative to the electrons as that of a directed beam
characterised by a single velocity, i.e. $V_{\rm max}\gg\vth^{\rm ion}$. From
the other hand, the analytical solution of Eq.~\eqref{eq:Silin} expressed in
Eq.~\eqref{eq:coll_freq} is only valid in the low-velocity limit $V_{\rm max} <
\vth$, where the friction coefficient $R$ is constant. The value of the maximum
relative velocity $V_{\rm max}$ depends both on the laser field strength and
$\nubar$ according to Eq.~\eqref{eq:V_max}.

The end result in Eq.~\eqref{eq:coll_freq} formally resembles the Gould-DeWitt
ansatz \cite{Gericke1996} due to the splitting of the collision frequency into a
sum of two contributions resulting from the strong and weak interactions.
However, here no {\em ad hoc} assumption was made and the splitting to the hard
and weak collision contributions in Eq.~\eqref{eq:coll_freq} was obtained as a
result of the discussed solution process. Thus, the present approach might
also hint on the region of applicability of the Gould-deWitt scheme when used
in other models.
\begin{figure}[t]
\includegraphics[width=8.6cm,clip=true]{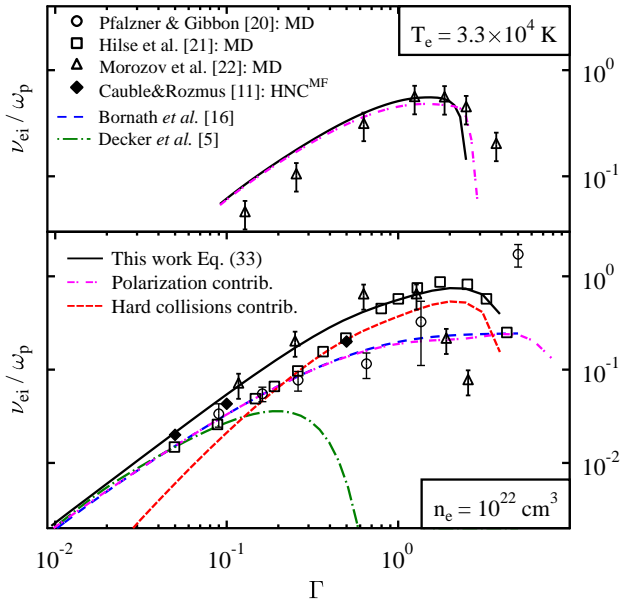}
\caption{Collision frequency $\nu_{ei}$ versus coupling parameter
         $\Gamma$ for a hydrogen plasma with fixed temperature / density
	 and a laser field with $\omega_0/\omega_p \!=\! 1$ (upper panel),
	 $\omega_0/\omega_p \!=\! 3$ (lower panel) and $v_0/\vth \!=\! 0.2$.
         Solid line: Eq.~\eqref{eq:coll_freq}; punctured lines:
	 contributions of hard collisions and polarisation to
	 Eq.~\eqref{eq:coll_freq}. The classical results of
	 Decker {\em et al.} \cite{Decker1994}) was calculated with
	 an integral cut-off at $k_{max} \!=\! m_e \vth^2 / Ze^2$.}
\label{fig:nu_vs_Gamma}
\end{figure}

The hard electron-ion collisions are incorporated in Eq.~\eqref{eq:coll_freq}
via a friction force related to the stopping power of the ions in an electron
gas  that in turn sets up a more general KH frame instead of the freely
oscillating one adopted in other works
\cite{Dawson1962,Decker1994,Kull2001,Bornath2001}. Many models have been
developed for the stopping power \cite{Zwicknagel1999}, few include hard
collisions. Within quantum statistical theory, they can be described by a
T-matrix approach based on the quantum Boltzmann equation. The related cross
sections are calculated from numerical solutions of the Schr\"odinger equation
\cite{Gericke1999}. The full stopping power can be then determined applying the
Gould-DeWitt scheme \cite{Gericke1996} or by velocity-dependent screening length
\cite{Gericke2002,Gericke2003}. According to the conditions of 
Fig.~\eqref{fig:nu_vs_Gamma}, the first approach is sufficient and has been used
to generate the data presented here. That is, the hard collision term has been
calculated as the stopping power using full cross sections (T-matrix approach)
minus the one in static first Born approximation.
\begin{figure}[t]
\includegraphics[width=8.6cm,clip=true]{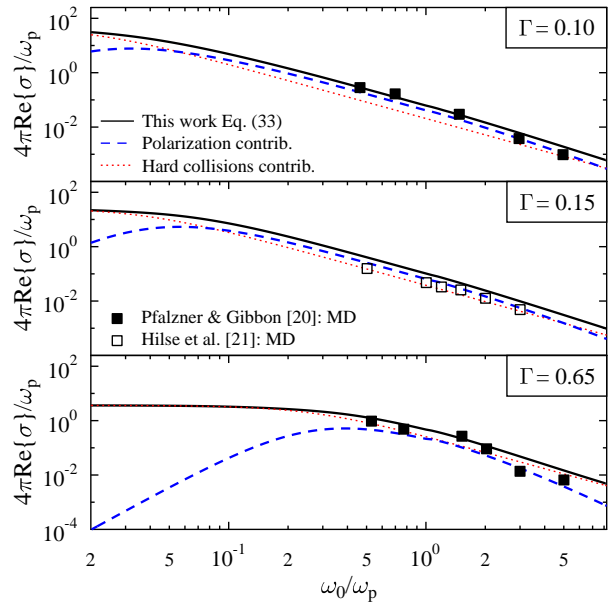}
\caption{Real part of the conductivity as a function of the laser frequency
	 for a hydrogen plasma $n_e = 10^{22}$cm$^{-3}$,
	 $v_0/\vth = 0.2$.}
\label{fig:nu_vs_omega}
\end{figure}

Next, the numerical solutions of Eq.~\eqref{eq:coll_freq} are compared
with the earlier approaches. We start with a commonly accepted benchmark
case first considered by \citeauthor{Dawson1962} and later used in works by
different authors. It is presented in Fig.~\ref{fig:nu_vs_Gamma} where our
results (e.g. Eq.\eqref{eq:coll_freq}) are compared with other theories
\cite{Bornath2001,Kull2001,Decker1994,Cauble1985} and simulation data
\cite{Hilse2005,Pfalzner1998,Morozov2005}. Yet another reason for choosing this
particular regime is that it presents the hardest parameter set for comparison
since it is lying inbetween the known limiting cases. As expected, all theories
agree for weakly coupled plasmas, but large deviations occur for strong
coupling. The classical description is clearly not applicable here as
demonstrated by comparison with the \citeauthor{Decker1994} result (also see
Ref.~\onlinecite{Grinenko2009} for extended discussion). For a coupling strength
of $\Gamma\gtrsim1$, the quantum theories of
Refs.~\onlinecite{Bornath2001,Kull2001} also start to disagree with the
simulation data. Such discrepancy can be traced back to the neglect of hard
collisions. Here this shortcoming is overcome. As a result we find an excellent
agreement with data from MD simulations by \citeauthor{Hilse2005} and
\citeauthor{Morozov2005}. Despite the inconsistency between these two data sets,
our result reproduces the main features observed in both simulations, especially
the sharp increase of $\nu_{ei}$ for $\Gamma\gtrsim 1$ relative to other
theories \cite{Kull2001,Bornath2001} and the change in the sign of the slope. 
Plotting both contributions of Eq.~\eqref{eq:coll_freq} separately reveals that
the hard-collisions term dominates for high coupling strengths and defines the
shape of the curve in this parameter area. From this, the observed change in
the slope sign is due to the turnover in the hard-collision contribution. The
latter occurs because the static Born overtakes the T-matrix contribution at
high coupling strengths.

Note, that different parameter sets are tested in the upper and lower frames of
Fig.~\ref{fig:nu_vs_Gamma}, the data running as a function of the increasing
density in the upper and decreasing temperature in the lower frame, and good
agreement between the MD results and our approach is obtained for this wide
data range. However, the contribution of hard-collisions is rather small for
the parameters presented in the upper frame, and therefore the advantages of
the current approach are less pronounced in that case. 

Degeneracy might obscure the comparison at low temperatures or high densities.
However, degeneracy is neither included in the MD simulations nor in our
calculation of the hard collision term which is based on solutions of
two-particle Schr\"odinger equation. We therefore compare our data to the MD
simulations on a similar level of approximation.

The multi-dimensional parameter space should also be examined along the
direction of the laser frequency. This comparison is demonstrated in
Fig.~\ref{fig:nu_vs_omega} for fixed electron density and the ratio of $v_0/\vth
= 0.2$, where the conductivity is plotted as a function of the laser frequency.
Here again we show the contributions of Eq.~\eqref{eq:coll_freq} separately. As
expected, the contribution of the dynamic, polarisation term vanishes at low
frequencies $\omega_0\ll\omega_p$, and the conductivity is dominated here by the
Drude-like term due to the hard-collisions. At the intermediate frequency range
$\omega_0 \sim \omega_p$, both terms contribute equally, with the dynamic term
overtaking for lower values of $\Gamma$. As discussed earlier, the present
approach is only valid for the laser frequencies that are of the same order of
magnitude as the plasma frequency and lower, therefore we do not extend the
comparison to the high laser frequencies. The breakdown of the present approach
at high frequencies is already exhibited in Eq.~\eqref{eq:coll_freq}. As follows
from the latter, at the high frequency limit the hard-collision dominate the
absorption, moreover $\nu_{ei} \xrightarrow{\omega_0\gg\omega_p} \nusc={\rm
const}$, which is obviously sensless. 

In conclusion, a quantum mechanical approach for the calculation of collisional
absorption of laser light in dense plasmas was presented. It consistently
incorporates the dynamic, field dependent response and hard electron-ion
collisions, in contrast to the earlier approaches that neglected the effect of
the latter \cite{Dawson1962,Decker1994,Kull2001,Bornath2001}. The use of
the quantum mechanical formulation allows avoiding the use of {\em ad hoc}
cutoffs and thus the theory remains reliable for strong electron-ion
interactions. The hard-collisions were introduced via the average friction force
due to the stopping power. Therefore, for the first time a link between
the stopping power and the problem of collisional laser absorption is
drawn. It allows applying the many theories developed for the stopping
power \cite{Zwicknagel1999} to the problem of collisional absorption. Although
only results for the quasi-linear regime $v_0/\vth \!\leq 1\!$ were presented,
the approach can be easily extended to higher field amplitudes, correlated
ions, and multiple ionisation stages.

The authors thank J.~Vorberger (CFSA, Warwick) for fruitful discussions and
EPSRC for financial support.


\end{document}